\RequirePackage{fix-cm}

\makeatletter
\def\@cons#1#2{\begingroup\let\@elt\relax\xdef#1{\ifx#1\relax\else#1\fi\@elt #2}\endgroup}
\makeatother

\documentclass[epj]{svjour}

\usepackage[utf8]{inputenc}
\usepackage[USenglish]{babel}
\usepackage{csquotes}
\usepackage{jenny}
\usepackage{graphicx}
\usepackage{hyperref}
\usepackage{url}
\usepackage{amsmath}
\usepackage[capitalise,noabbrev]{cleveref}
\usepackage{subcaption}
\usepackage{float}
\usepackage{cite}
\usepackage{color, colortbl}
\usepackage{xcolor}
\usepackage{multirow}
\usepackage{mathtools}
\usepackage{siunitx}
\usepackage{soul}

\usepackage{glossaries}

\newacronym{pwa}{PWA}{Partial Wave Analysis}
\newacronym{cm}{c.m.}{center-of-mass}
\newacronym{QN}{QN}{quantum numbers}
\newacronym{pawian}{PAWIAN}{PArtial Wave Interactive ANalysis}
\newacronym{NLL}{NLL}{Negative LogLikelihood}
\newacronym{BIC}{BIC}{Bayesian Information Criterion}
\newacronym{AIC}{AIC}{Aikake Information Criterion}
\newacronym{KL}{KL}{Kullback-Leibler}

\newacronym{DTF}{DTF}{DecayTreeFitter}
\newacronym{FAIR}{FAIR}{Facility for Antiproton and Ion Research}
\newacronym{HESR}{HESR}{High Energy Storage Ring}

\definecolor{LightGreen}{HTML}{CDF3CB}
\definecolor{LightOrange}{HTML}{FFD580}

\begin{document}\sloppy
	\raggedbottom
	\hugehead
	%

\title{Prospects for Spin-Parity Determination of Excited Baryons via the $\mathbf{\bar{\Xi}^+\Lambda\,K^-}$ Final State with PANDA}

%
%
\author{
V.~Abazov\inst{16} \and 
V.~Abramov\inst{54} \and 
P.~Achenbach\inst{37} \and 
A.~Akram\inst{65} \and 
M.~Al-Turany\inst{15} \and 
M.~Albrecht\inst{4} \and 
G.~Alexeev\inst{16} \and 
A.~Ali\inst{20} \and 
W.~Alkakhi\inst{4} \and 
C.~Amsler\inst{67}
% ...
}
\institute{
Università Politecnica delle Marche-Ancona,{ \bf Ancona}, Italy \and 
Universität Basel,{ \bf Basel}, Switzerland \and 
Institute of High Energy Physics, Chinese Academy of Sciences,{ \bf Beijing}, China \and 
Ruhr-Universität Bochum, Institut für Experimentalphysik I,{ \bf Bochum}, Germany \and 
Bolu Abant Izzet Baysal University,{ \bf Bolu}, Turkey \and 
Rheinische Friedrich-Wilhelms-Universität Bonn,{ \bf Bonn}, Germany \and 
Università di Brescia,{ \bf Brescia}, Italy \and 
Institutul National de C\&D pentru Fizica si Inginerie Nucleara "Horia Hulubei",{ \bf Bukarest-Magurele}, Romania \and 
Chiang Mai University,{ \bf Chiang Mai}, Thailand \and 
University of Technology, Institute of Applied Informatics,{ \bf Cracow}, Poland \and 
IFJ, Institute of Nuclear Physics PAN,{ \bf Cracow}, Poland \and 
AGH, University of Science and Technology,{ \bf Cracow}, Poland \and 
Instytut Fizyki, Uniwersytet Jagiellonski,{ \bf Cracow}, Poland \and 
FAIR, Facility for Antiproton and Ion Research in Europe,{ \bf Darmstadt}, Germany \and 
GSI Helmholtzzentrum für Schwerionenforschung GmbH,{ \bf Darmstadt}, Germany \and 
Joint Institute for Nuclear Research,{ \bf Dubna}, Russia \and 
University of Edinburgh,{ \bf Edinburgh}, United Kingdom \and 
Friedrich-Alexander-Universität Erlangen-Nürnberg,{ \bf Erlangen}, Germany \and 
Northwestern University,{ \bf Evanston}, U.S.A. \and 
Goethe-Universität, Institut für Kernphysik,{ \bf Frankfurt}, Germany \and 
INFN Laboratori Nazionali di Frascati,{ \bf Frascati}, Italy \and 
Dept of Physics, University of Genova and INFN-Genova,{ \bf Genova}, Italy \and 
Justus-Liebig-Universität Gießen II. Physikalisches Institut,{ \bf Gießen}, Germany \and 
IRFU, CEA, Université Paris-Saclay,{ \bf Gif-sur-Yvette Cedex}, France \and 
University of Glasgow,{ \bf Glasgow}, United Kingdom \and 
KVI-Center for Advanced Radiation Technology (CART), University of Groningen,{ \bf Groningen}, Netherlands \and 
Gauhati University, Physics Department,{ \bf Guwahati}, India \and 
University of Science and Technology of China,{ \bf Hefei}, China \and 
Universität Heidelberg,{ \bf Heidelberg}, Germany \and 
Department of Physics, Dogus University,{ \bf Istanbul}, Turkey \and 
Istanbul Okan University, { \bf Istanbul }, Turkey \and
Forschungszentrum Jülich, Institut für Kernphysik,{ \bf Jülich}, Germany \and 
Karlsruhe Institute of Technology, Institute for Data Processing and Electronics,{ \bf Karlsruhe}, Germany \and 
Chinese Academy of Science, Institute of Modern Physics,{ \bf Lanzhou}, China \and 
INFN Laboratori Nazionali di Legnaro,{ \bf Legnaro}, Italy \and 
Lunds Universitet, Department of Physics,{ \bf Lund}, Sweden \and 
Johannes Gutenberg-Universität, Institut für Kernphysik,{ \bf Mainz}, Germany \and 
Helmholtz-Institut Mainz,{ \bf Mainz}, Germany \and 
Research Institute for Nuclear Problems, Belarus State University,{ \bf Minsk}, Belarus \and 
Institute for Theoretical and Experimental Physics named by A.I. Alikhanov of National Research Centre "Kurchatov Institute”,{ \bf Moscow}, Russia \and 
Moscow Power Engineering Institute,{ \bf Moscow}, Russia \and 
Moscow Engineering Physics Institute, National Research Nuclear University, {\bf Moscow}, Russia \and 
Westfälische Wilhelms-Universität Münster,{ \bf Münster}, Germany \and 
Suranaree University of Technology,{ \bf Nakhon Ratchasima}, Thailand \and 
Synchrotron Light Research Institute,{ \bf Nakhon Ratchasima}, Thailand \and 
Nankai University,{ \bf Nankai}, China \and 
Novosibirsk State University,{ \bf Novosibirsk}, Russia \and 
Budker Institute of Nuclear Physics,{ \bf Novosibirsk}, Russia \and 
University of Wisconsin Oshkosh,{ \bf Oshkosh}, U.S.A. \and 
Dipartimento di Fisica, Università di Pavia, INFN Sezione di Pavia,{ \bf Pavia}, Italy \and 
University of West Bohemia,{ \bf Pilsen}, Czech \and 
Charles University, Faculty of Mathematics and Physics,{ \bf Prague}, Czech Republic \and 
Czech Technical University, Faculty of Nuclear Sciences and Physical Engineering,{ \bf Prague}, Czech Republic \and 
A.A. Logunov Institute for High Energy Physics of the National Research Centre “Kurchatov Institute”,{ \bf Protvino}, Russia \and 
National Research Centre "Kurchatov Institute" B. P. Konstantinov Petersburg Nuclear Physics Institute, Gatchina,{ \bf St. Petersburg}, Russia \and 
Stockholms Universitet,{ \bf Stockholm}, Sweden \and 
Kungliga Tekniska Högskolan,{ \bf Stockholm}, Sweden \and 
Sardar Vallabhbhai National Institute of Technology, Applied Physics Department,{ \bf Surat}, India \and 
Veer Narmad South Gujarat University, Department of Physics,{ \bf Surat}, India \and 
Florida State University,{ \bf Tallahassee}, U.S.A. \and 
Department of Physics, University of Tehran, North Karegar Avenue { \bf Tehran}, Iran \and
INFN Sezione di Torino,{ \bf Torino}, Italy \and 
Università di Torino and INFN Sezione di Torino,{ \bf Torino}, Italy \and 
Politecnico di Torino and INFN Sezione di Torino,{ \bf Torino}, Italy \and 
Uppsala Universitet, Institutionen för fysik och astronomi,{ \bf Uppsala}, Sweden \and 
National Centre for Nuclear Research,{ \bf Warsaw}, Poland \and 
Österreichische Akademie der Wissenschaften, Stefan Meyer Institut für Subatomare Physik,{ \bf Wien}, Austria 
}

\date{\today}



\abstract{
	A study of the baryon excitation spectra provides a deep insight into the inner structure of baryons.
	Most of the world-wide efforts have been directed towards $N^*$ and $\Delta^*$ spectroscopy.
	Complementary data from double and triple strange baryon spectra are lacking and foreseen to be obtained with the \panda experiment in the near future.
	Earlier Monte Carlo studies demonstrated that with an expected cross section in the order of $\mu$b, \panda will be able to copiously observe the channel \mychannelfs, including the two resonances \excitedcascadesixteen and \excitedcascadetwenty, with a negligible background contribution.
	In this study, the feasibility to determine the spin and parity of the \excitedcascadesixteen and  \excitedcascadetwenty resonances is investigated by making use of a partial wave analysis employing the PAWIAN software.
	The presented results demonstrate the capability of the \panda experiment to determine the spin-parity of these resonances with a few days of data taking.
}
\maketitle
\newpage
\setcounter{page}{4}

\glsdisablehyper

\section{Introduction}
	The strength of the strong interaction depends on the strong coupling constant $\alpha_s$, which increases with decreasing momentum transfer.
At a length scale corresponding to the proton radius the value of $\alpha_s$ is so large that perturbative calculation methods no longer are applicable.
Certain regions of this non-perturbative regime are still not well investigated, for example the multi-strangeness sector. 
Furthermore, there are still open questions to be answered, e.g. what are the relevant degrees-of-freedom of baryons? Is there a three-quark or quark-di-quark structure?
In the low energy regime, the exchange of hadrons describes the appropriate degrees of freedom for the scattering cross section of baryonic resonances. 
Since hadrons, i.e. baryons and mesons, are according to the quark model composite particles, they have internal degrees of freedom and thus an excitation pattern.
To gain a deeper insight into the mechanisms of non-perturbative QCD, it is essential to understand the excitation pattern of baryons.
There are theoretical models, which are used to predict hadronic processes in this kinematic regime.
And they need to be constraint by experimental data.
Two classes of theoretical approaches are currently well established: Lattice Quantum Chromodynamics (LQCD) \cite{Wilson1974} which solves the non-perturbative QCD by using numerical simulations, and effective theories that exploit the chiral symmetry of the QCD Lagrangian \cite{gasser1985,Weinberg1990,epelbaum2009,petschauer2020,hyodo2020}.
LQCD has given impressive results for hadron spectroscopy \cite{Lin2008,Horsley2011,Padmanath2018} and low-energy physics \cite{aoki2014,della2020,illa2020,aoki2020} during the last decades.\\
There are two possibilities to study hadrons in experiments.
The first one, is to study reaction dynamics, i.e. the investigation of hadron-hadron interactions and hadron production to study ground state properties.
The other is hadron spectroscopy, where the structure of hadrons is investigated.
In this work the focus is on hadron spectroscopy.\\
Both methodologies require an interplay between theory and experiment.
During the last decades various calculations of the baryon spectrum within the Constituent Quark Model (CQM) \cite{capstick1986,glozman1998,lring2001} have been performed.
The CQM describes the (anti)baryon as a system consisting of three (anti)quarks, which are bound by a confining interaction.
Most experimental studies have focused on the nucleon excitation spectrum.
In contrast, the knowledge of excited double or triple strange baryon states, called hyperons, is poor.
According to the SU(3) symmetry, the $\Xi$ spectrum, for example, has as many states as the $N^*$ and $\Delta$ spectrum together.
Since hyperons are unstable particles, they unveil more information on their characteristics than stable nucleons.
Therefore, hyperon decays are a powerful tool to address physics challenges like the fundamental symmetries and the internal structure of baryons.
Furthermore, the study of the excited double strange system will validate the information deduced from the nucleon spectrum.\\
The excitation spectra of most hyperons as well as their ground state properties are still not well understood.
These reactions provide a good opportunity to gain access these properties and spectra, since a high fraction of the inelastic \pbarp cross section is associated to final states with a baryon-antibaryon pair together with additional mesons.
Furthermore, it is possible to directly populate intermediate states, where one or both hyperons are in an excited state.
These excited states will give rise to final states consisting of an antibaryon-baryon pair and one or more mesons.
The produced particles may further decay weakly or electromagnetically.
If the resonant states in the (anti-)baryon-meson combined system are sufficiently narrow, it will be possible to reconstruct the invariant mass from the final state products.
A \gls{pwa} will then give the opportunity to determine complementary information about the properties, e.g. spin and parity quantum numbers, which are difficult to determine directly.\\
Next generation experiments are planned to perform comprehensive studies of the strangeness baryon spectrum.
For instance, Jefferson Lab recently approved the proposal to deploy a $K_L$ beam \cite{Amaryan2020}.
The facility will be able to produce around $5.3\cdot 10^6$ \excitedcascadetwenty events within 100 days of beam on target.
Furthermore, the future Antiproton Annihilation in Darmstadt (\panda) experiment located at the FAIR facility  \cite{FAIR2019} will be well-suited for a comprehensive baryon spectroscopy program in the multi-strange sector \cite{PhaseOnePaper,Erni2009}.
\panda will be a multi-purpose detector to study antiproton-proton induced reactions at beam energies between $1.5\momentumunit$ and $15\momentumunit$.
The production cross sections are expected to be on the order $\mu\mt{b}$ \cite{Musgrave1965} for final states consisting exclusively of a \anticascade\cascade pair.
This gives the possibility to produce $10^6$ (excited) \cascade events per day at an assume luminosity of $\mathcal{L}+10^{31} \mt{cm}^{-2}\mt{s}^{-1}$.\\
A feasibility study for the reconstruction of the reaction \mychannel and its \cc channel with the \panda detector, where $\Xi^*$ denotes the following intermediate resonances: \excitedcascadesixteen and \excitedcascadetwenty, has been performed \cite{Puetz2020}.
The study showed that \panda is able to reconstruct between \SI{3}{\percent} and \SI{5}{\percent} of the generated events with good signal significance to analyze this channel and sufficient background suppression.
The data sample measured with \panda will be subject to a \gls{pwa} of the \fs final state.
A \gls{pwa} allows to extract complex amplitudes of certain processes from experimental data with the aim to investigate its dynamics.
If a process is dominated by resonances, the \gls{pwa} gives the possibility to determine their masses and widths as well as their spins and \gls{QN}.
In addition, it is possible to determine the coupling strengths to the production and decay processes.
\panda exclusively measures \pbarp-annihilations \enquote{in-flight} meaning that the total momentum of the \pbarp system is not zero.
The production of the \fs final state requires a \gls{cm} energy of $\sqrt{s}\geq\SI{2.93}{\GeV}$, which corresponds to an antiproton momentum of $p_{\bar p}\geq\SI{3.5}{\GeV\per c}$.\\
In this study, the feasibility of a PWA of the \fs final state, including specific $\Xi$ resonances in the \lam\kminus system, is investigated with the PArtial Wave Interactive ANalysis (PAWIAN) software \cite{Kopf2019}
with focus on the spin and parity determination.
A brief description of the software framework is given in \cref{sec:software}.
An overview on how the \gls{pwa} is carried out, i.e. which amplitudes have been used and which assumptions have been made is given in \cref{sec:description}. 
This is followed by the event generation in \cref{sec:evtGen}. 
\Cref{sec:SelectionCriteria} introduces the selection criteria used to compare the fit results.
The investigations to determine the spin and parity QN of the specific $\Xi$ resonances are presented in \cref{sec:pwa}.
\Cref{sec:summary} summarizes the results and provides an outlook.

	\label{sec:introduction}
	
\section{\pandabf Detector}
	\label{sec:panda}
\begin{figure*}[htb]
		\centering
		\includegraphics[width=0.85\textwidth]{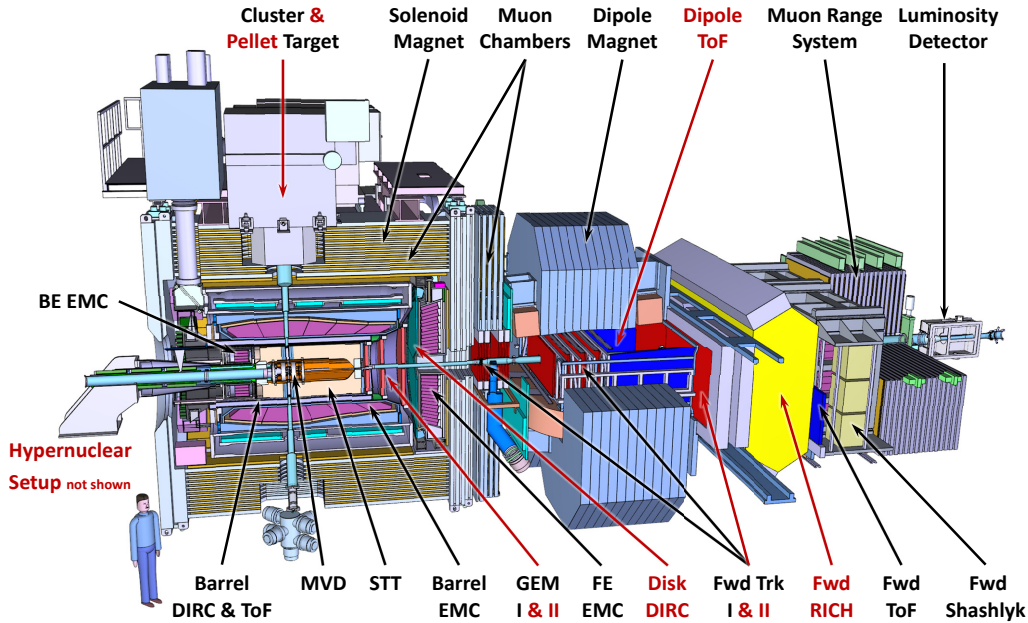}
		\caption{Schematic overview of the \panda detector setup. The components with black labels will be available for the initial configuration of \panda and the components with red labels will be added in a later stage of the experiment. Figure taken from \cite{PANDADetector}.}
		\label{fig:PANDADetector}
	\end{figure*}

	The \panda detector \cite{PhaseOnePaper}, shown in \cref{fig:PANDADetector}, is a multi-purpose detector and will be an internal experiment at the \gls{HESR} which is one of the storage rings at FAIR.
The \gls{HESR} is optimized for high energy antiprotons and will provide a luminosity of $\mathcal{L}=10^{31}\mt{cm}^{-2}\mt{s}^{-1}$ in the first phase and a peak luminosity of $\mathcal{L}=10^{32}\mt{cm}^{-2}\mt{s}^{-1}$ in a later stage of operation.
The detector will be composed of two parts: the \textit{Target Spectrometer} (TS) surrounding the interaction point (IP) and the \textit{Forward Spectrometer} (FS).
At \panda interactions between the antiproton beam and fixed target protons and/or nuclei will be investigated at \gls{cm} energies between $2.25\unit{GeV}$ and $5.47\unit{GeV}$.
The target protons will be provided either by a cluster-jet or frozen hydrogen pellets \cite{PandaTDRTarget}.
Due to its modular design, \panda will provide a nearly complete angular coverage as well as high resolutions for charged and neutral particles and a good particle identification (PID).\\
The Micro Vertex Detector (MVD) is the innermost part of the tracking system inside the TS and uses two different detector technologies: hybrid pixel detectors and double-sided micro-strip detectors \cite{PandaTDRMVD}.
Its main task is to tag events with open charm and strangeness, to provide high precision on the determination of the position information and angular distribution and to improve the precision of the transverse momentum.\\
The MVD is surrounded by the Straw Tube Tracker (STT), which consists of 4224 single straw tubes arranged in a cylindrical volume around the IP \cite{PandaTDRSTT}.
MVD, STT and the Gaseous Electron Multiplier (GEM) planes, which are downstream of the STT, are embedded inside the magnetic field of a $2\,\mt{T}$ solenoid \cite{PandaTDRMagnets} giving the possibility to measure the momentum of charged particles. \\
In the FS, the main tracking system for charged particles is called the Forward Tracker (FTrk) and consists of three pairs of tracking planes equipped with straw tubes \cite{PandaTDRFTS}.
The main task of the FTrk is to measure particles with low transverse momentum.
Therefore, the tracking planes will be placed before, inside and behind a $2\unit{T}\cdot\mt{m}$ dipole magnet.\\
For the event reconstruction a good PID is important.
Therefore, the \panda design includes various PID sub-detectors, i.e. Cherenkov detectors, in particular the Detection of Internal Cherenkov Light (DIRC) \cite{PandaTDRDirc} and the Ring Imaging Cherenkov (RICH) detector, the Barrel Time of Flight (BarrelToF) \cite{PandaTDRBarrelToF} and the Forward Time of Flight (FToF) detector \cite{PandaTDRFToF} as well as Muon Detector System (MDS) \cite{PandaTDRMDS}.\\
The Electromagnetic Calorimeter (EMC) will provide an efficient reconstruction of scattering angles and momenta of positrons, electrons, and photons while the background will be suppressed efficiently.
In the TS the EMC consists of three sub-detectors, ie. the Backward-Endcap EMC (BE EMC), the Barrel EMC and the Forward-Endcap EMC (FE EMC), and it will be equipped with more than 15,000 PbW$\mt{O}_4$ crystals \cite{PandaTDREMC}.
A shashlyk-type calorimeter is foreseen \cite{PandaTDRFWEMC} in the FS.
The FS will be completed with a Luminosity Detector (LMD) to enable cross section normalization by measuring forward elastically scattered antiprotons in the Coulomb-Nuclear interference region \cite{PandaTDRLumi}.

\section{Software}
	\label{sec:software}
	\subsection{PandaRoot}
The software framework used to simulate, reconstruct and analyze the data is called PandaRoot \cite{Pandaroot} and is based on ROOT \cite{Brun1997} together with the Virtual Monte Carlo (VMC) package \cite{Hrivnacova2003}.
The simulation and reconstruction code is implemented within the FairRoot software framework \cite{Al-Turany2012}, which is developed as a common computing structure for the future FAIR experiments.
The detector simulation is handled by VMC giving the possibility to use Geant3 \cite{brun1993geant} or, used in this study, Geant4 \cite{Agostinelli2003}.
The detector response, including the digitization, after the simulation and propagation of the events is simulated as well.
In the reconstruction, charged particle tracks are formed by combining the hits from the tracking detectors.
To take magnetic field inhomogeneities, energy loss, small angle scattering and the error calculation for the different detector parts into account, the Kalman Filter GENFIT \cite{Rauch2015} and the track follower GEANE \cite{Fontana2008} are used.\\
In the recent version of PandaRoot the tracking algorithms use the IP as the origin of the particle track.
As a consequence, they do not perform well for the reconstruction of hyperons, which decay with displaced vertices due to their relative long lifetime and thus far from the IP.
Hence, an ideal tracking algorithm, which groups the hit points into a track based on the generated particle information, is used for the reconstruction of hyperons.
The influence of using ideal tracking on the reconstruction of hyperons has been discussed in \cite{Puetz2021}.
Based on this discussion, no impact of the simplification on the conclusion of this work is expected.\\
For PID, the information of the PID detectors are correlated to the information coming from the reconstructed charged particles tracks.
Neutral candidates are formed from clusters inside the EMC for which no correlated tracks were found.
For a fast PID, algorithms based on Bayesian approaches are implemented \cite{Pandaroot}.

\subsection{PAWIAN}
The PAWIAN software package has been developed at the Ruhr University Bochum under the GNU General Public License \cite{Kopf2014}.
It is mainly designed for analyses dedicated to the \panda physics program but also supports other experiments in the field of hadron spectroscopy such as BESIII \cite{BESIII} and CrystalBarrel@LEAR \cite{Albrecht2020,Amsler2015}.
One of the important software features is the possibility to use different reactions, i.e. \pbarp  annihilation and $e^+e^-$ annihilation, spin formalism and dynamic functions \cite{Pychy2016}.
PAWIAN provides an event-based maximum-likehood fit with different minimizers and an option for parallel processing.
In addition, PAWIAN provides the option of a single channel or coupled channel analysis.
In this work, we only consider a single channel analysis and thus only the relevant configuration parts for a single channel analysis are described.
Further information can be found in \cite{Pychy2016} and \cite{Pawian2020}.\\
The PAWIAN package contains an event generator which is used in this study to generated the data samples to be analyzed.
Within the event generation the four-momenta of the particles in the data set are generated based on quantum mechanical rules.
The experimental or generated data are then subject to the minimization process.
The outcome of this fit procedure can be used to extract physical quantities like pole positions, coupling strengths or spin density matrix elements.

\setstcolor{red}
\section{Partial Wave Analysis}
	\label{sec:description}
	The inelastic reaction of an antiproton with a proton is a strong process and thus all \gls{QN} of the initial state are conserved.
In addition, the initial state defines the basis for the description of the total amplitudes.
\begin{table}[b]
	\centering
	\caption{Antiproton-proton initial states up to $J=6$. The allowed quantum numbers for the intermediate states are presented in the form $^{\left(2S+1\right)}L_J$. Adapted from \cite{Puetz2020}.}
	\label{tab:QNinitialState}
	\begin{tabular}{ccccccc}
		\hline
		$J$ & Singlet & $J^{PC}$ & Triplet & $J^{PC}$ & Triplet & $J^{PC}$ \\
		& $\lambda=0$ & & $\lambda=\pm 1$ & & $\lambda=0,\pm1$\\
		\hline\hline
		0 & $^1S_0$ & $0^{-+}$ & & & $^3P_0$ & $0^{++}$ \\
		1 & $^1P_1$ & $1^{+-}$ & $^3P_1$ & $1^{++}$ & $^3S_1,\,^3D_1$ & $1^{--}$ \\
		2 & $^1D_2$ & $2^{-+}$ & $^3D_2$ & $2^{--}$ & $^3P_2,\,^3F_2$ & $2^{++}$ \\
		3 & $^1F_3$ & $3^{+-}$ & $^3F_3$ & $3^{++}$ & $^3D_3,\,^3G_3$ & $3^{--}$\\
		4 & $^1G_4$ & $4^{-+}$ & $^3G_4$ & $4^{--}$ & $^3F_4,\,^3H_4$ & $4^{++}$\\
		5 & $^1H_5$ & $5^{+-}$ & $^3H_5$ & $5^{++}$ & $^3G_5,\,^3I_5$ & $5^{--}$ \\
		6 & $^1I_6$ & $6^{-+}$ & $^3I_6$ & $6^{--}$ & $^3H_6,\,^3J_6$ & $6^{++}$\\
		\hline	
	\end{tabular}
\end{table}
The possible QN combinations for the antiproton-proton initial states up to $J=6$ are listed in \cref{tab:QNinitialState}.
For the description of this part of the partial wave amplitude, the formation of the $J^{PC}$ has to be taken into account.
Otherwise, the elements of the spin density matrices would not fulfill the required symmetry.
\begin{figure}[t]
	\centering
	\includegraphics[width=0.4\textwidth]{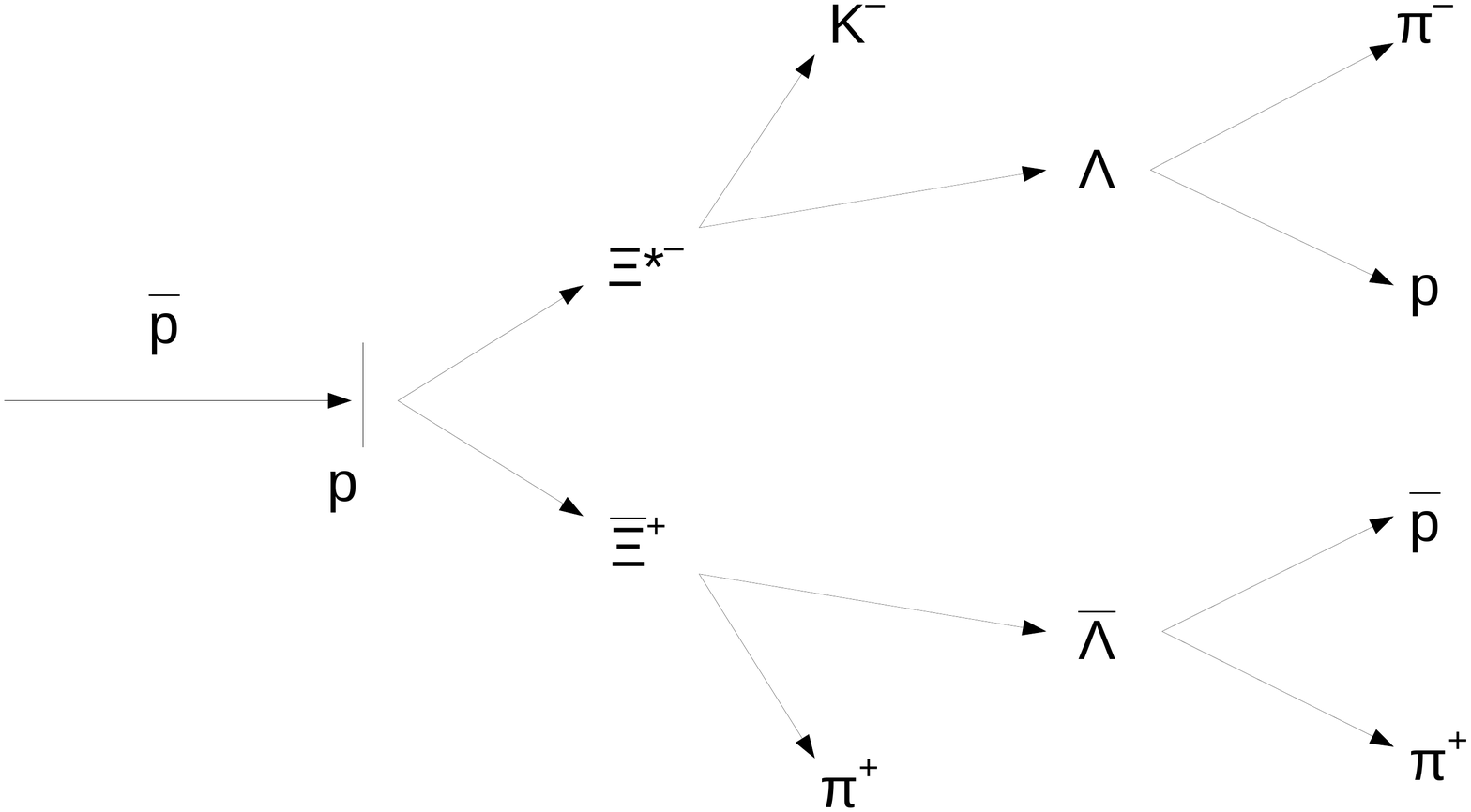}
	\caption{Illustration of the reaction chain including the hyperon weak decay.}
	\label{fig:DecayTree}
\end{figure}
The decay chain, shown in \cref{fig:DecayTree}, starting from the initial \pbarp states over the produced excited $\Xi$ resonances down  \anticascade, \lam and \kminus, are described by the helicity formalism.
As simplification, in this study the weak decay of \anticascade and \lam is not taken into account.
However, it is expected that decay asymmetry of the hyperon weak decay will set additional constraints.
Details on the reconstruction of the decay chain can be found in \cite{Puetz2021}.
To get excess to orbital momentum dependent barrier factors the helicity amplitudes are further expanded into the $LS$-scheme.
The details about these amplitudes used in the study here can be found in \cite{Albrecht2020}.\\
The antiproton momentum defines the maximum angular momentum $L_{\max}$ and thus the number of contributing amplitudes.
A good approximation for $L_{\max}$ is given by truncating the expression \cite{Pychy2016}:
\begin{equation}
	L_{\max}\approx \frac{p_{\bar{p}}^{\mt{cm}}}{200\unit{MeV/c}},
	\label{eq:Lmax}
\end{equation}
where $p_{\bar{p}}^{\mt{cm}}$ denotes the antiproton momentum in the rest frame of the \pbarp system.
In case of an endothermic reactions, the maximum orbital momentum is determined by the mass of the final state particles.\\
For the fit of the \pbarp channels an unbinned maximum likelihood minimization method is used and also described in detail in \cite{Albrecht2020}.

\section{Event Generation}
	\label{sec:evtGen}
	Within the PAWIAN framework, there is the possibility to generate events based on a user-defined decay model or based on fit results obtained with real data.
In this study the first approach is used to generate events for the reaction \mychannelfs.
As already mentioned in \cref{sec:introduction}, the production of the \fs final state requires an antiproton momentum of $p_{\bar{p}}\geq 3.5\momentumunit$.
In this study the antiproton momentum is chosen to be $4.6\momentumunit$, corresponding to a c.m. energy of $\sqrt{s} =3.25\unit{GeV}$.
This energy is about $300\unit{MeV}$ above the production threshold of the final state particles and allows the population of the two resonances, \excitedcascadesixteen and \excitedcascadetwenty, that are the main subject in this study, but not sufficient for the production of two excited $\Xi$ states in one event.\\
As described in \cref{sec:description}, in endothermic reactions the maximum orbital momentum is determnined by the mass of the final state particles, here the mass of the resonances and the \anticascade.
The chosen beam momentum implies a momentum in the \gls{cm} frame of $p^\mt{cm} \approx 600\unit{MeV/c}$ for \excitedcascadesixteen and $p^\mt{cm} \approx 410\unit{MeV/c}$ for \excitedcascadetwenty.
According to \cref{eq:Lmax}, the maximum orbital angular momentum is $L_{\max} = 3$ for \excitedcascadesixteen and $L_{\max} = 2$ for \excitedcascadetwenty.
Beside $L_{\max}$ certain parameters are required for the event generation, namely, the mass and width of the resonances as well as the magnitude and phase of all possible initial LS combinations for the \pbarp system.
The dynamics are modeled by a a Breit-Wigner function.\\
Various data sets are needed for the study of the final state.
Since the normalization of the fit uses Monte-Carlo-Integration, the first data set consists of phase-space distributed events called the continuum data set in the following. The other data set, called the signal data set, includes events generated according to the following reactions:
\begin{enumerate}
	\item[i.] \pbarp $\rightarrow \bar{\Xi}^+$\excitedcascadesixteen or \pbarp $\rightarrow \bar{\Xi}^+$\excitedcascadetwenty
	\item[ii.] \pbarp $\rightarrow \bar{\Xi}^+$\excitedcascadesixteen and \pbarp $\rightarrow \bar{\Xi}^+$\excitedcascadetwenty
\end{enumerate}
In the first case, only one of the resonances is generated, while in the latter case both resonances are included in the signal data set.
To perform a PWA of a three-body final state, statistics between 1,000 and 100,000 events are used \cite{ablikim2013,ablikim2013b,munzer2018}.
The results obtained in \cite{Puetz2020} indicated that about 30,000 events are needed to perform the PWA with PAWIAN.
Based on a \SI{5}{\percent} reconstruction efficiency obtained in \cite{Puetz2021} for the reconstruction of the \fs final state for the initial configuration of \panda, a data sample of about 600,000 generated events is needed to ensure that a sufficient number of reconstructed events are available for the fit after passing the full PandaRoot simulation and reconstruction chain.
In addition, the continuum data set has to pass the PandaRoot simulation and reconstruction criteria to ensure that the data sample underlies the same detector acceptance as the signal. 
Therefore, about 1.8 million phase-space distributed events have been generated.
The properties of the generated data sets are summarized in 
\cref{tab:EvtGen}.
Furthermore a set of parameters according to the assumption for the spin and parity \gls{QN} is needed for the event generation.
\begin{table}[t]
	\centering
	\caption{Properties of the generated continuum and signal data sets.}
	\label{tab:EvtGen}
	\begin{tabular}{llrr}
		\hline
		\multicolumn{2}{c}{Reaction}  & $\#$Events & $L_{\max}$ \\
		\hline \hline
		\multirow{2}{*}{continuum}&\mychannelfs & 1,800,000 & 2 \\
		& \mychannelfs & 1,800,000 & 3 \\
		\hline
		\multirow{2}{*}{signal (i.)} &\pbarp $\rightarrow\bar{\Xi}^+$\excitedcascadesixteen & 600,000 & 3 \\
		& \pbarp $\rightarrow\bar{\Xi}^+$\excitedcascadetwenty & 600,000 & 2 \\
		\hline		
		\multirow{2}{*}{signal (ii.)}&\pbarp $\rightarrow\bar{\Xi}^+$\excitedcascadesixteen + & \multirow{2}{*}{600,000} & \multirow{2}{*}{3} \\
		& \pbarp $\rightarrow\bar{\Xi}^+$\excitedcascadetwenty & & \\
		\hline
	\end{tabular}
\end{table}

\section{Selection Criteria}
	\label{sec:SelectionCriteria}
	To determine the spin and parity QN as well as the mass and width of the resonances, it is necessary to specify the correct model among a set of tested models.
Therefore, it is essential to compare the consistency of the data and the fit with the complexity of the used model.
Since the fit is performed with different models, a comparison of the resulting negative loglikelihood (NLL) values is not possible and thus different criteria have to be used to compare the models.
In this study, the Aikake Information Criterion (AIC) and Bayesian Information Criterion (BIC) are used.\\
The AIC \cite{Burnham2004} is based on 
the so-called Kullback-Leibler (KL) divergence that describes the information loss if model $B$ is used as an approximation for model $A$.
In general, model $A$ is unknown and the KL divergence cannot be used directly.
Therefore, a relative KL divergence for a different model $B$ is determined by
\begin{equation}
	\mt{AIC} = -2 \ln \left(\mathcal{L}\right) + 2K,
\end{equation}
where $\mathcal{L}$ is the maximized likelihood and $K$ the number of free parameters.
This definition implies that among a set of models, the model with the smallest AIC value is preferred as best model.
However, the single AIC is not interpretable, since it contains arbitrary constants.
Therefore, a rescale to \cite{Burnham2004}
\begin{equation}
	\Delta_i = \mt{AIC}_i - \mt{AIC}_{\min},
\end{equation}
with labeling hypothesis number $i$ is used to rank the candidate hypotheses.
According to this, the preferred model has $\Delta=0$, while all other models will have $\Delta>0$.
Commonly, the relative merit is assessed by the following rules:
	\begin{itemize}
		\item[$\bullet$] $\Delta_i\leq 2$: model is supported;
		\item[$\bullet$] $4\leq \Delta_i \leq 7$: model has less support;
		\item[$\bullet$] $\Delta_i>10$: model is not supported.
	\end{itemize}
These rules are based on the rules used in the Bayesian literature \cite{Raftery1995}.
According to this, the confidence level for the rejection of a model with $\Delta_i>10$ is more than $95\%$.
Another selection criterion is the BIC which has been developed \cite{Schwarz1978} and is defined by
\begin{equation}
	\mt{BIC} = - 2\ln\left(\mathcal{L}\right) + K \log\left(n\right),
\end{equation}
where $n$ denotes the number of events.
It is based on the likelihood function and closely related to the AIC.
Similar to AIC, the model with the smallest BIC value is the most preferable one.
Since the BIC is asymptotically consistent, for an infinite number of data points, the true model will be selected, if it is included in the set of tested models.
The sole usage of the BIC can lead to underfitting meaning that not always the true model is selected.
In contrast, the sole usage of AIC can lead to overfitting since it does not depend on the sample size.
This leads to two possible scenarios: BIC and AIC prefer the same model or they prefer different models.
In the first case the best hypothesis can be selected by using the $\Delta$AIC criterion while in the latter case good experience has been made with a model selection based on the sum of AIC and BIC \cite{Pychy2016}.
In this case, the model with the smallest value for $S_i = \mt{AIC}_i + \mt{BIC}_i$ is preferred and the model selection is done by comparing the $\Delta$(AIC$+$BIC).

\section{Results of the Partial Wave Analysis}
	\label{sec:pwa}
Systematic studies for the \gls{pwa} of the \fs final state have been performed to investigate various scenarios, i.e. a single resonance case as well as including a crossed channel.
The aim of these studies is to provide information about the feasibility to determine the \gls{QN} of the final state with PAWIAN as well as the effects of additional resonances in the \anticascade\kminus system \cite{Puetz2020}.
In particular the aim is to study the feasibility to determine the \gls{QN} from \panda data.

\subsection{Single Resonance Case}
	\label{sec:systematics}
\begin{figure}[t]
	\centering
	\begin{subfigure}[t]{0.9\linewidth}
		\includegraphics[width=0.8\linewidth]{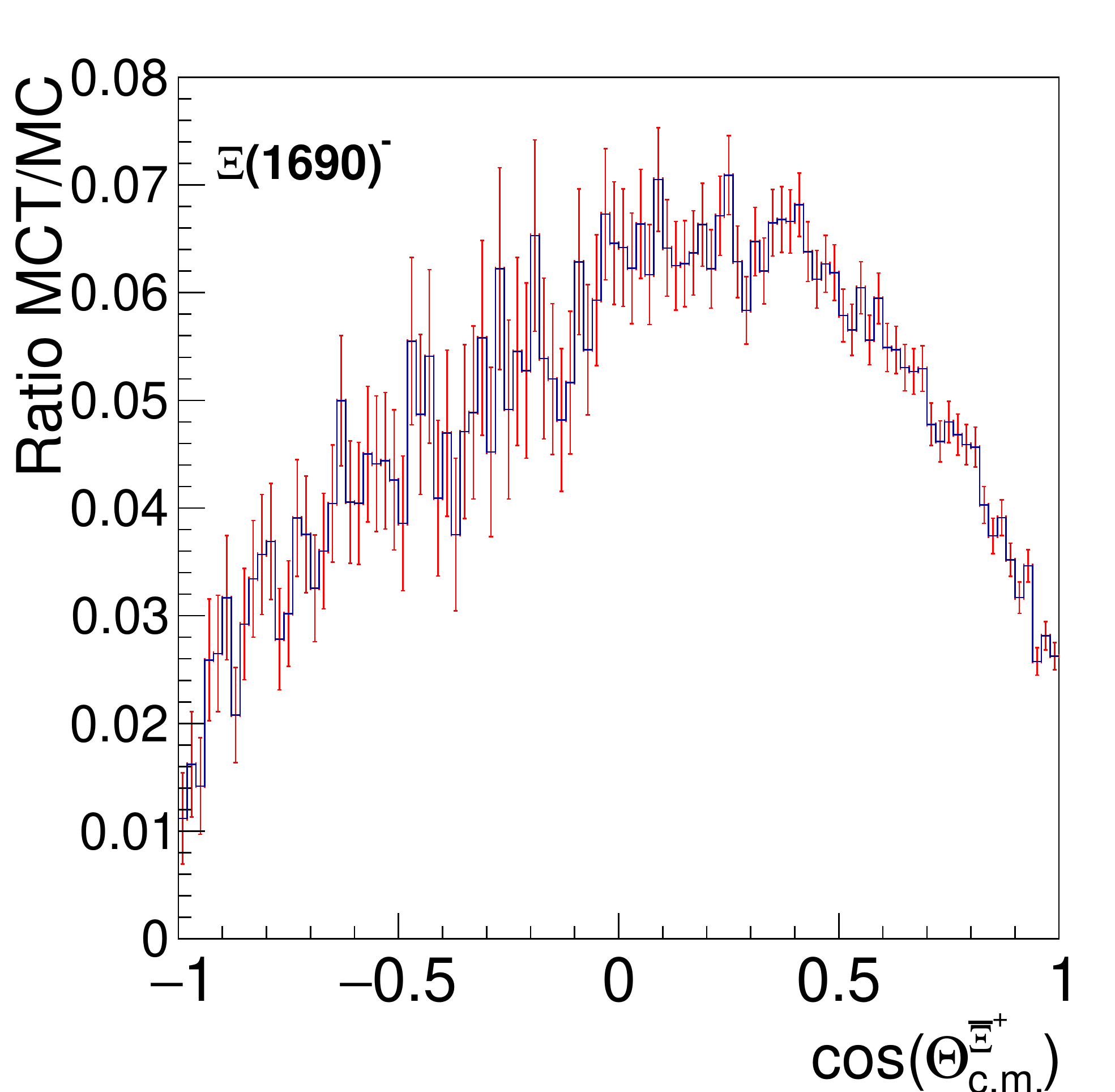}
		\caption{}
	\end{subfigure}
	\begin{subfigure}[t]{0.9\linewidth}
		\includegraphics[width=0.8\linewidth]{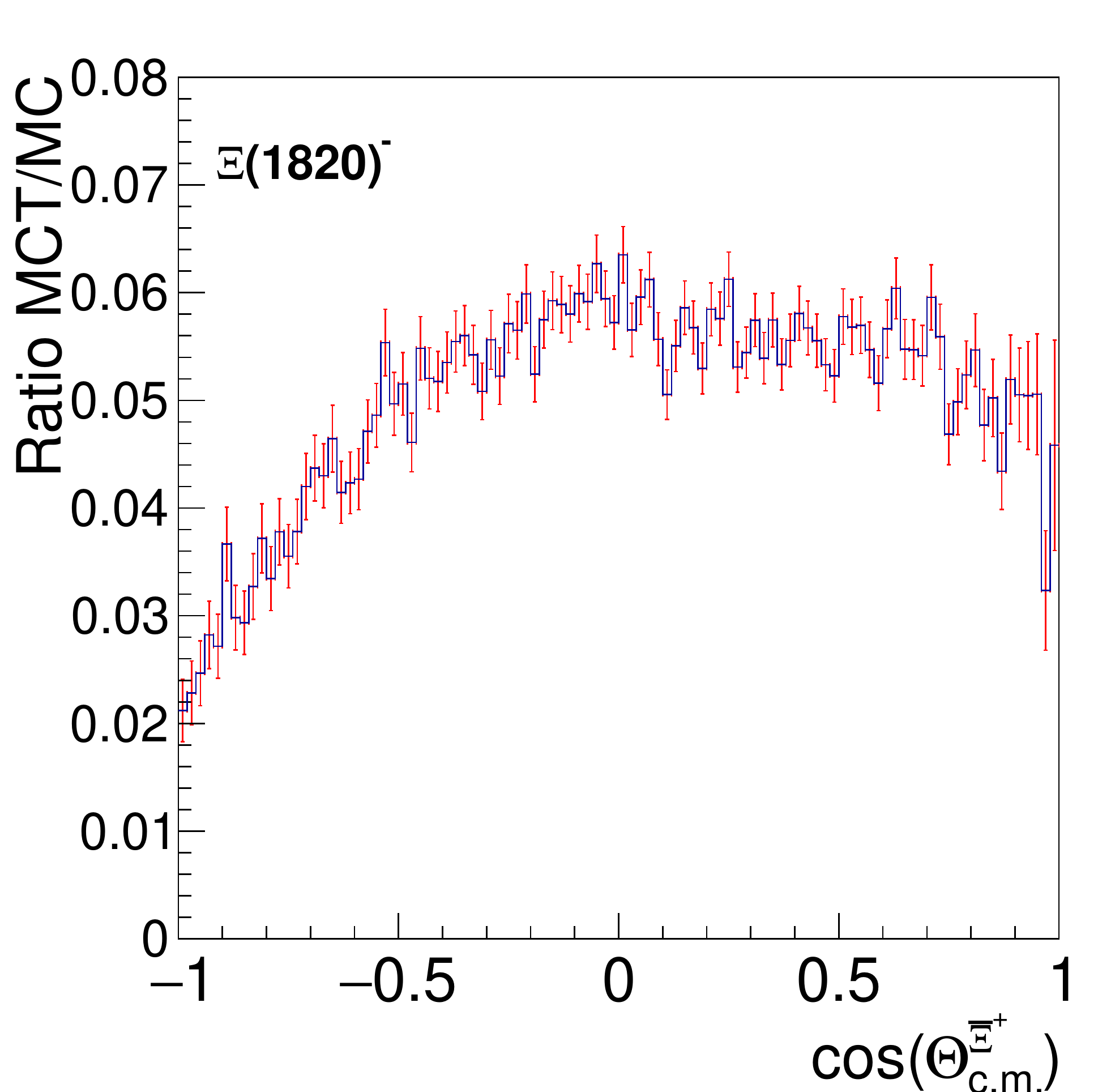}
		\caption{}
	\end{subfigure}
	\caption{Ratio of \costheta distributions for the Monte-Carlo truth partners of the final reconstructed (MCT) and the generated (MC) \anticascade candidates in the \acrshort{cm} frame. a) shows the ratio for the \excitedcascadesixteen sample with generated $J^P=1/2^+$ hypothesis and b) for the \excitedcascadetwenty sample with generated $J^P=3/2^-$ hypothesis.}
	\label{fig:Xi1690_SingleRes_Hyp1p_CosThetaRatio_Xibar}
\end{figure}

\begin{figure}[htbp]
	\centering
	\includegraphics[width=0.85\linewidth]{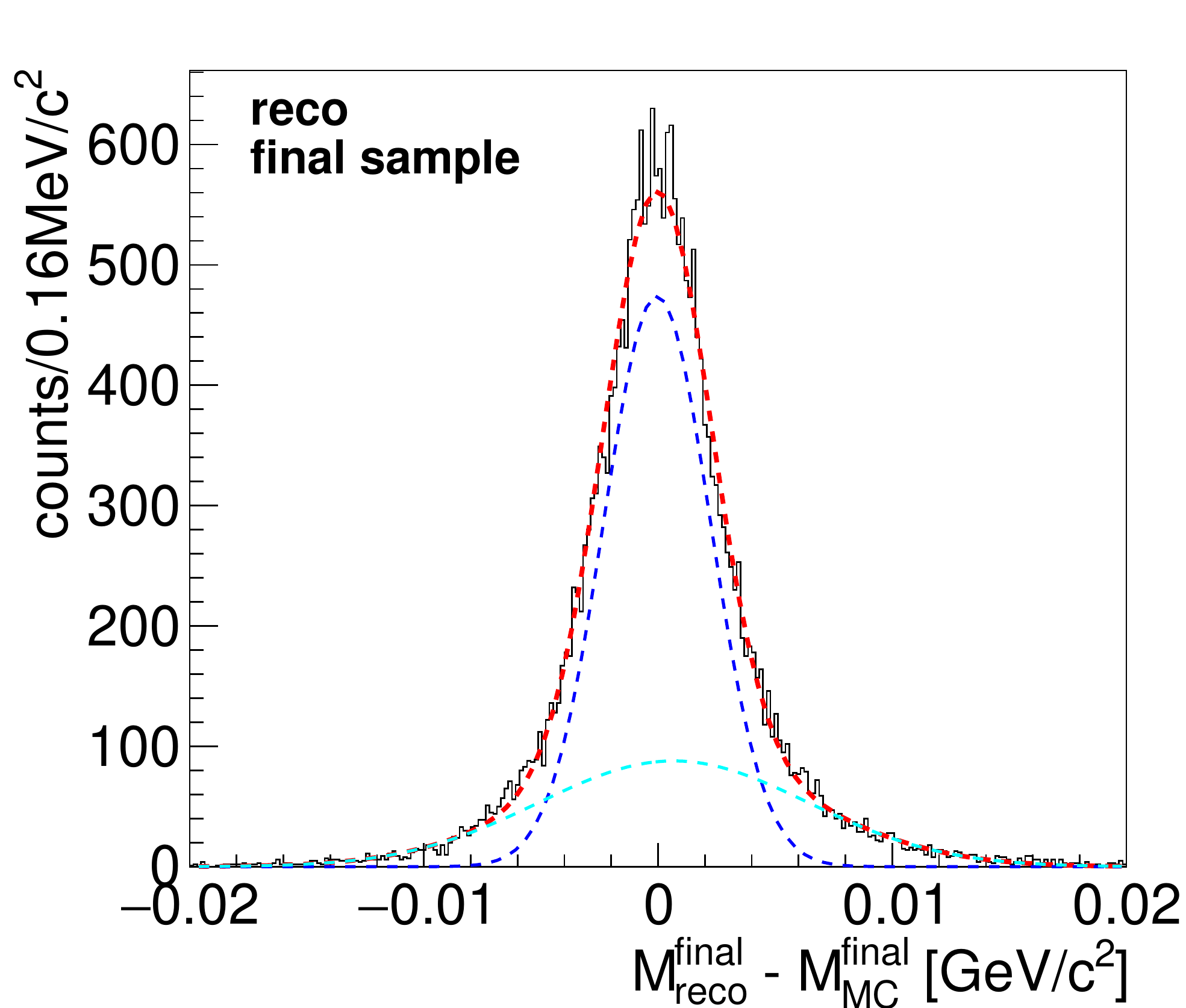}
	\caption{Mass resolution of the reconstructed \lam\kminus system (black histogram). The distribution is fitted with a double Gaussian function (red dashed line). The inner Gaussian function is shown as blue dashed line and the second Gaussian function is indicated by the cyan dashed line.}
	\label{fig:Xi1690_SingleRes_LK_dm}
\end{figure}

	To investigate the influence of the \panda detector acceptance, the resolution as well as the full reconstruction efficiency on the fit results, we start by presenting a simplified study of the single resonance case.
The respective signal data samples have been generated with the event generator included in PAWIAN and were then used as input for the simulation and reconstruction chain in PandaRoot.
After running the PandaRoot simulation and reconstruction, the remaining events were analyzed, employing a combined kinematic and geometric fit (\gls{DTF})\cite{Puetz2021}.
Analogous to the signal data sets, the continuum data set has been subject to the analysis procedure.\\
For each resonance, data sets for each of the different spin and parity QN, i.e. $1/2^+$, $1/2^-$, $3/2^+$ and $3/2^-$ have been generated.
About \SI{5}{\percent} to \SI{6}{\percent} of the generated events passed the full simulation and reconstruction chain in PandaRoot and were used as candidates for the fit.
This is in agreement with the reconstruction efficiency obtained in \cite{Puetz2021}.
The reconstruction efficiency for a generated $J=3/2$ hypothesis is slightly higher than the one for a $J=1/2$ hypothesis.
For a large part this can be explained by more reconstructed events in the region between \costhetacm$=0.5$ and \costhetacm$=1$ for the $3/2$ hypothesis, while for the $1/2$ hypothesis a loss of efficiency is observable in this angular region.
$\Theta_\mt{c.m.}$ is defined as the angle between the beam axis and the momentum vector of the reconstructed particle in the \gls{cm} frame. 
The higher efficiency is also indicated in \cref{fig:Xi1690_SingleRes_Hyp1p_CosThetaRatio_Xibar} showing the ratios for the $\cos\Theta_\mt{c.m.}$ distribution of the generated partners (MCT) of the reconstructed \anticascade and all generated candidates in the \gls{cm} frame.
The illustration shows that no holes in the acceptance are observable in the full \costheta range, which holds for all generated hypotheses.\\
In addition to the reconstruction efficiency and the detector acceptance, the mass resolution for the single $\Xi$ resonances has been investigated.
Therefore, the deviation of the reconstructed from the generated mass has been fitted with a double Gaussian function, as illustrated in \cref{fig:Xi1690_SingleRes_LK_dm}.
Since the width of the inner gaussian function seems to underestimate the detector resolution, the Root Mean Square (RMS) of the distribution is used to estimate the mass resolution.
The mass resolutions are summarized in \cref{tab:MassResolution_SingleRes}.
\begin{table}[b]
	\centering
	\caption{Mass resolution for the $\Xi$ resonances from the different reconstructed samples.}
	\label{tab:MassResolution_SingleRes}
	\begin{tabular}{lccc}
		\hline
		Resonance & Hypothesis & inner Gauss& RMS\\
		& & [$\si{\mega\electronvolt\per c\squared}$] & [$\si{\mega\electronvolt\per c\squared}$] \\
		\hline
		\hline
		\multirow{4}{*}{\excitedcascadesixteen} & $1/2^+$ & $2.3$ & $3.5$\\
		 & $1/2^-$ & $2.2$ &  $3.6$ \\
		 & $3/2^+$ & $2.2$ & $3.4$\\
		 & $3/2^-$ & $2.2$  & $3.5$\\
		 \hline
 		\multirow{4}{*}{\excitedcascadetwenty} & $1/2^+$ & $2.3$ & $4.0$\\
		 & $1/2^-$ & $2.2$ & $3.8$\\
		 & $3/2^+$ & $ 2.0$ & $3.5$\\
		 & $3/2^-$ & $2.0$ & $3.9$\\
		 \hline
	\end{tabular}
\end{table}
The obtained values are in agreement with the fit values determined in \cite{Puetz2020}.\\
30,000 events from each analysis output data sets have been subject to a multi-dimensional fit with PAWIAN.
In all tested cases the best fit result is achieved by using the true hypothesis for the multi-dimensional fit. 
The fit results for the resonances are summarized in \cref{tab:FullSimXi1690DeltaAIC,tab:FullSimXi1820AICBIC}.
For the \excitedcascadesixteen all wrong hypotheses can be excluded with high significance since $\Delta$AIC$>316$.
For the \excitedcascadetwenty resonance the model selection is based on AIC+BIC 
because the hypothesis with the smallest $\Delta$AIC does not match the one with the smallest $\Delta$BIC.
The difference between the sum for the second best and the best fit is in all cases greater than 96 indicating that all wrong hypotheses are not supported by the selection criteria.
\begin{table}[b]
	\centering
	\caption{$\Delta$AIC values for the data sample simulating the \excitedcascadesixteen resonances obtained from the multi-dimensional fit. The best fit results are marked in green.}
	\label{tab:FullSimXi1690DeltaAIC}
	\begin{tabular}{cc|rrrr}
		\hline
		& Fit $\rightarrow$  & $1/2^+$ & $1/2^-$ & $3/2^+$ & $3/2^-$ \\
		Gen. $\downarrow$ &  & & & &\\
		\hline
		\multicolumn{2}{l|}{$1/2^+$} &\cellcolor{LightGreen} 0.0 &	2,550.6 & 2,310.6& 	2,706.8\\
		\multicolumn{2}{l|}{$1/2^-$} & 316.7 &	\cellcolor{LightGreen} 0.0 & 328.2 & 2,332.2\\
		\multicolumn{2}{l|}{$3/2^+$} & 4,973.9 & 5,228.0 & 	\cellcolor{LightGreen} 0.0 &	584.6\\
		\multicolumn{2}{l|}{$3/2^-$} & 5,345.6 & 3,118.6 &	833.1 &	\cellcolor{LightGreen} 0.0\\
		\hline
	\end{tabular}
\end{table}
\begin{table}[htbp]
	\centering
	\caption{$\Delta$(AIC+BIC) values for \excitedcascadetwenty obtained from the multi-dimensional fit. The best fit results are marked in green.}
	\label{tab:FullSimXi1820AICBIC}
	\begin{tabular}{cc|rrrr}
		\hline
		& Fit $\rightarrow$  & $1/2^+$ & $1/2^-$ & $3/2^+$ & $3/2^-$ \\
		Gen. $\downarrow$ &  & & & &\\
		\hline
		\multicolumn{2}{l|}{$1/2^+$} &\cellcolor{LightGreen}  0.0 & 139.9 & 158.7 & 208.1\\
		\multicolumn{2}{l|}{$1/2^-$} & 96.8 & \cellcolor{LightGreen} 	0.0 & 211.1 & 887.4\\
		\multicolumn{2}{l|}{$3/2^+$} & 7473.3 & 7604.5 &	\cellcolor{LightGreen} 	0.0 & 198.4\\
		\multicolumn{2}{l|}{$3/2^-$} & 7617.6 & 6900.8 & 490.2 & \cellcolor{LightGreen} 0.0 \\
		\hline
	\end{tabular}
\end{table}
Since the relevant breakup momenta in the process \pbarp$\rightarrow$\anticascade\excitedcascade are expected to be small, it could be assumed that the amplitudes with large orbital momenta are strongly suppressed due to the barrier factors.
A study of the previous data samples has been performed, assuming that the maximum angular momentum of the resonances is larger with $L=3$.
This assumption is analogue to that for the initial \pbarp states.
The performed study showed that neither the number of free fit parameters was reduced significantly nor that the significance of the fit results increases.

\subsection{Two Resonances Case}
\begin{figure}[t]
	\centering
	\includegraphics[width=0.7\linewidth]{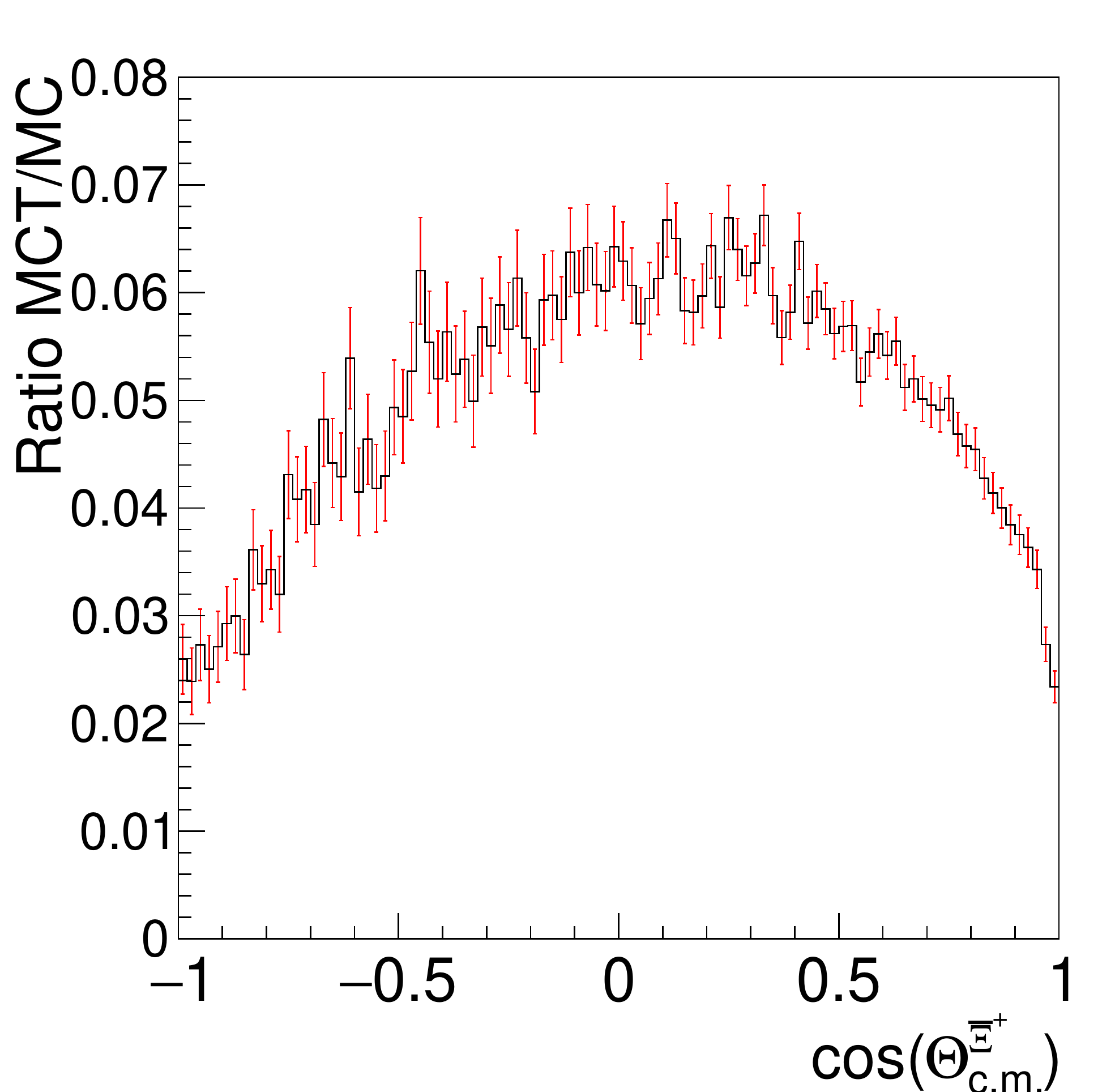}
	\caption{Ratio of \costhetacm distributions for the Monte-Carlo truth partners of the final reconstructed (MCT) and the generated (MC) \anticascade candidates in the \acrshort{cm} frame.}
	\label{fig:RatioCosThetaCMSCominedSample}	
\end{figure}

\begin{figure}[t]
	\centering
	\includegraphics[width=0.8\linewidth]{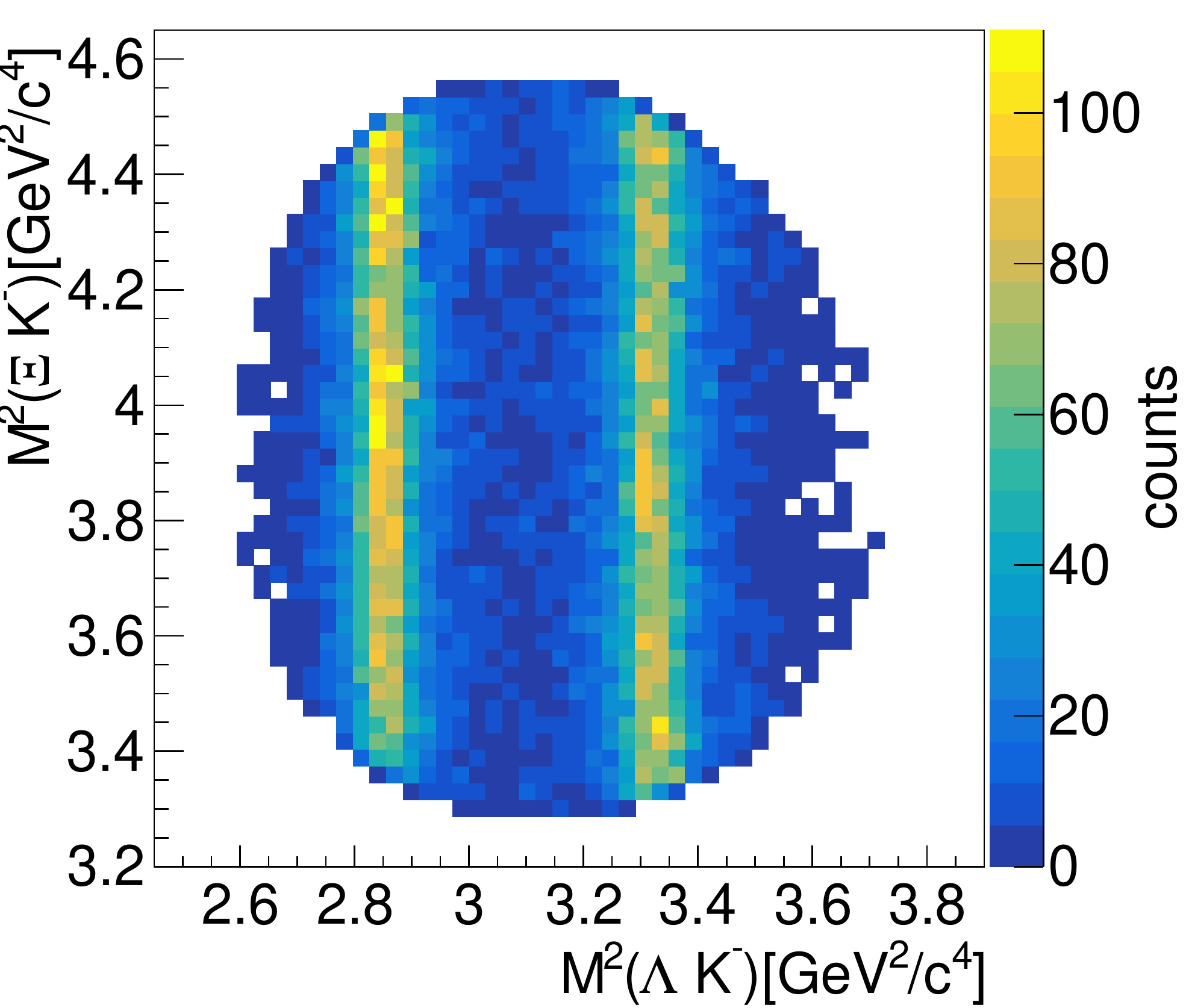}
	\caption{Dalitz plot of the final reconstructed sample.}
	\label{fig:CombinedSampleDalitzReco}
\end{figure}

\begin{figure}[t]
	\centering
	\includegraphics[width=0.8\linewidth]{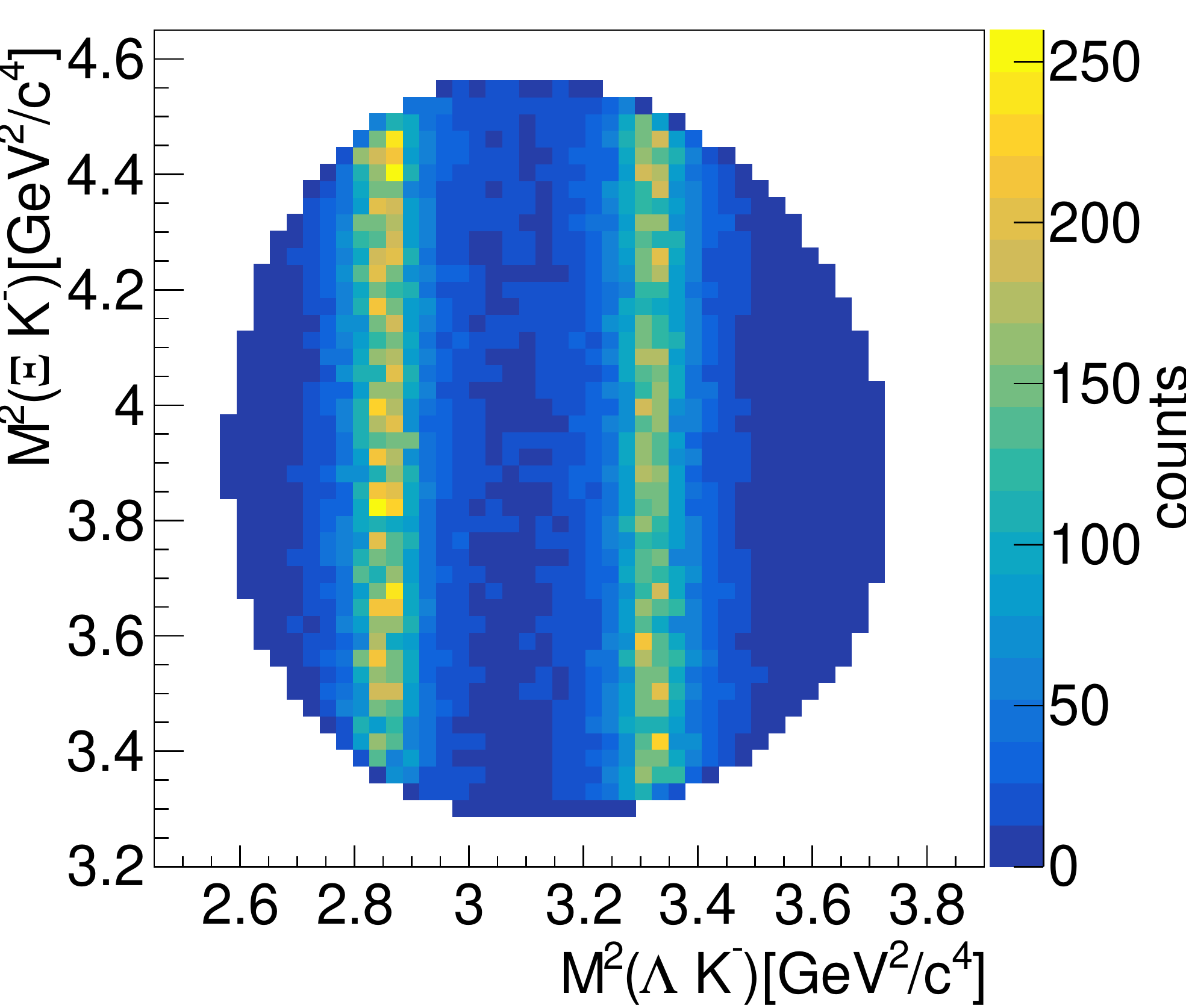}
	\caption{Dalitz plot of the fitted $1/2^+ 3/2^-$ sample.}
	\label{fig:CombinedSampleDalitzFit}
\end{figure}

\begin{figure}[b]
	\centering
	\includegraphics[width=0.8\linewidth]{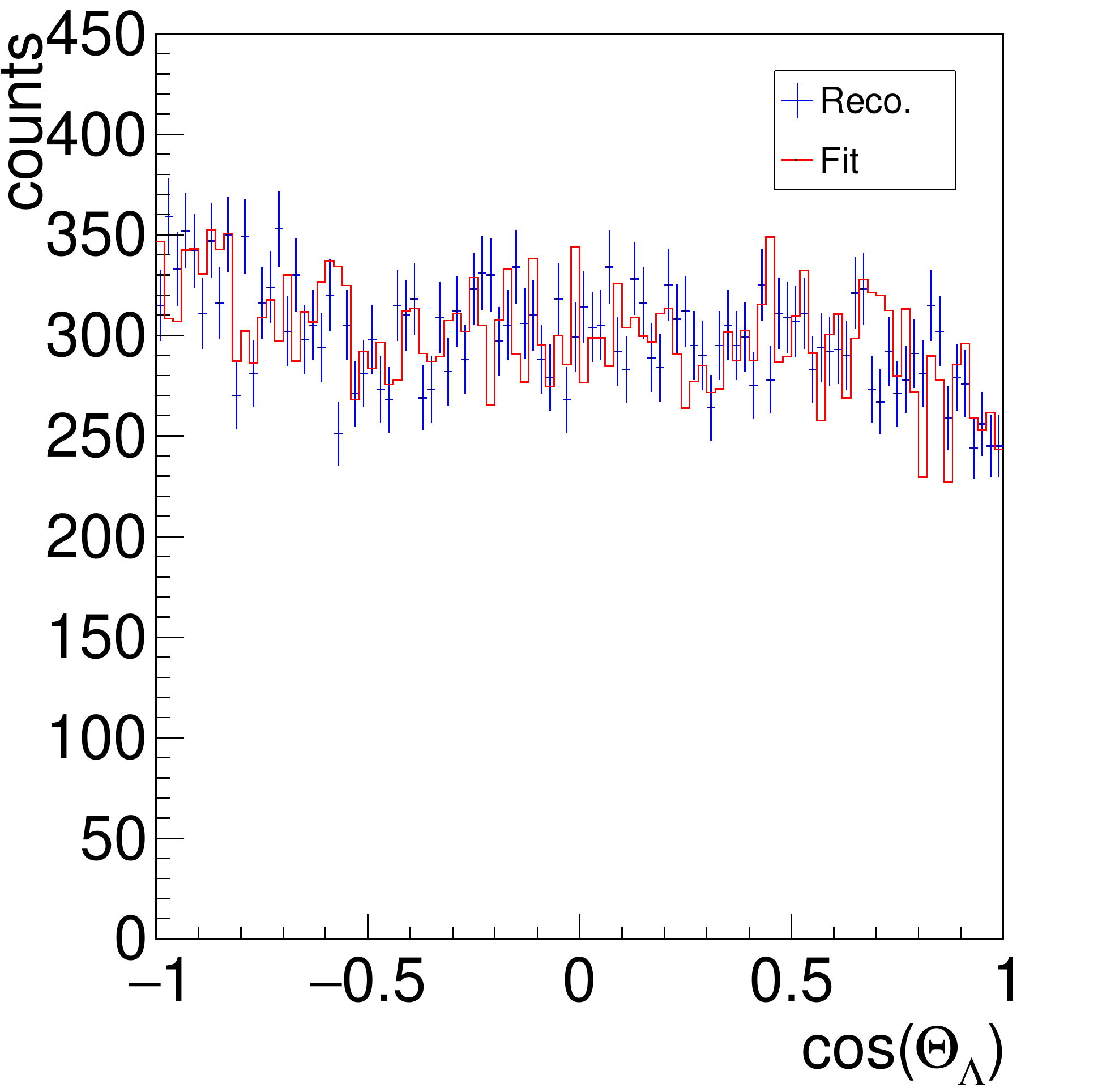}
	\caption{Reconstructed (Reco.) and fitted angular distribution for \lam in the helicity frame of the \excitedcascade for generated and fitted $1/2^+3/2^-$ hypothesis.}
	\label{fig:AngularDistrGenRecoLamHeli}
\end{figure}

	In the previous study, we discussed the feasibility to extract the proper \gls{QN} in the case one specific $\Xi$ resonance was considered.
In the following, we extend this study by considering the case in which two resonances, namely the \excitedcascadesixteen and \excitedcascadetwenty, are present.\\
For this study, a data set of 600,000 signal events has been generated using a $J^P=1/2^+$ hypothesis for \excitedcascadesixteen and $J^P=3/2^-$ for \excitedcascadetwenty, equating to 16 multi-dimensional fits four the QN hypotheses $1/2^+$, $1/2^-$, $3/2^+$, and $3/2^-$ for both resonances, respectively.
Since nothing is known about the production of the $\Xi$ resonances, the contribution of the resonances was assumed to be equal, meaning that the sample consists of \SI{50}{\percent} \mbox{\anticascade\excitedcascadesixteen} events and \SI{50}{\percent} \mbox{\anticascade\excitedcascadetwenty} events.
The generated data sample was subsequently used as input for the full simulation, reconstruction and analysis chain in PandaRoot.
\SI{5.1}{\percent} of the generated events have been reconstructed, whereof 30,000 events were used for the multi-dimensional fit procedure with PAWIAN.
The pure signal fraction of the data sample is about \SI{94}{\percent}, meaning that about \SI{6}{\percent} of the reconstructed sample consists of wrongly combined events passing the selection criteria.
\cref{fig:RatioCosThetaCMSCominedSample} shows the ratio of the \costhetacm distributions for the MCT partners of the final reconstructed and the generated \anticascade candidates in the \acrshort{cm} frame.
The figure shows that the reconstruction efficiency is between \SI{5}{\percent} and \SI{7}{\percent} within \costhetacm$=-0.5$ and \costhetacm$=0.5$. 
Furthermore, a loss of efficiency for the forward and backward direction is observable that is caused by the loss of particles in the beam pipe and other material.\\
\Cref{fig:CombinedSampleDalitzReco} shows the Dalitz plot of the reconstructed sample.
Both $\Xi$ resonances are identifiable as vertical structures.
The wrongly combined events appear in the region of the resonances and thus in a region with high statistics. 
A fit of the sample omitting the wrongly combined events showed that the amount of \textquote{wrong} events has no significant influence on the fit results.
The resulting data sample has then been subject to the 16 fits where the
hypotheses are combined by the four QN and are presented in the scheme $J^P\left(\Xi\left(1690\right)\right) J^P\left(\Xi\left(1820\right)\right)$ in the following.
Depending on the chosen hypothesis, the number of free fit parameters varies from 94 to 162.
\begin{table}[b]
	\centering
	\caption{Fit results for the combined sample with generated $1/2^+ 3/2^-$ hypothesis. The best fit is marked in green and the second best fit in orange.}
	\label{tab:CombinedSampleAICBIC}
	\begin{tabular}{cc|rrrr}
		\hline
		Fit. Hyp. &  & NLL & BIC & AIC & $N_{par}$\\
		\hline
		\multicolumn{2}{l|}{$1/2^+1/2^+$} & -6,605.8& -12,242.5& -13,023.6 & 94 \\
		\multicolumn{2}{l|}{$1/2^+1/2^-$}  &-22,587.7& 	-44,206.3& -44,987.3 & 94\\
		\multicolumn{2}{l|}{$1/2^+3/2^+$}  &-22,687.6 &	-44,055.7 & -45,119.2& 128\\
		\multicolumn{2}{l|}{\cellcolor{LightGreen}$1/2^+3/2^-$}  &\cellcolor{LightGreen}-22,888.0 &\cellcolor{LightGreen}	-44,456.4 &\cellcolor{LightGreen} -45,519.9&\cellcolor{LightGreen} 128\\		
		\multicolumn{2}{l|}{$1/2^-1/2^+$}  &-22,552.9 &	-44,136.8 & -44,917.8& 94\\
		\multicolumn{2}{l|}{$1/2^-1/2^-$}  & -22,363.7& -43,758.4 & -44,539.4 & 94\\
		\multicolumn{2}{l|}{$1/2^-3/2^+$}  &-9,811.8 &-18,304.1 & -19,367.7& 128 \\
		\multicolumn{2}{l|}{$1/2^-3/2^-$}  &-9,068.9 &	-16,818.3 & -17,881.9 & 128\\
		\multicolumn{2}{l|}{$3/2^+1/2^+$} &	-22,581.1 &	-43,842.7 & -44,906.3 & 128 \\
		\multicolumn{2}{l|}{$3/2^+1/2^-$}  &-22,614.7 &-43,909.9 & -44,973.5 & 128\\
		\multicolumn{2}{l|}{$3/2^+3/2^+$}  &  -22,691.0& -43,712.7 & -45,058.8& 162\\
		\multicolumn{2}{l|}{\cellcolor{LightOrange}$3/2^+3/2^-$}  &\cellcolor{LightOrange}-22,844.0 &\cellcolor{LightOrange}-44,018.0 &\cellcolor{LightOrange} -45,364.0 &\cellcolor{LightOrange} 162\\
		\multicolumn{2}{l|}{$3/2^-1/2^+$} &	-22,513.2 	&-43,706.9 & 	-44,770.5 & 128 \\
		\multicolumn{2}{l|}{$3/2^-1/2^-$}  &-22,483.0 &-43,646.4 & 	-44,709.9 & 128\\
		\multicolumn{2}{l|}{$3/2^-3/2^+$}  &-22,705.9 &-43,741.7 & 	-45,087.7 & 162\\
		\multicolumn{2}{l|}{$3/2^-3/2^-$}  & -22,656.3 &-43,642.5 & 	-44,988.6& 162\\
		\hline
	\end{tabular}
\end{table}
The results of the fits are summarized in \cref{tab:CombinedSampleAICBIC}.
The difference of the AIC value between the second best (marked in orange) and the best fit (marked in green) is $\Delta\mt{AIC} > 150$.\\
\cref{fig:CombinedSampleDalitzFit} shows the Dalitz plot deduced from the best fit, namely $1/2^+ 3/2^-$.
The comparison of the Dalitz plot of the fitted and the final reconstructed sample 
used as input for the \gls{pwa} shows that both are in agreement.
\begin{figure}[t]
	\centering
	\includegraphics[width=0.8\linewidth]{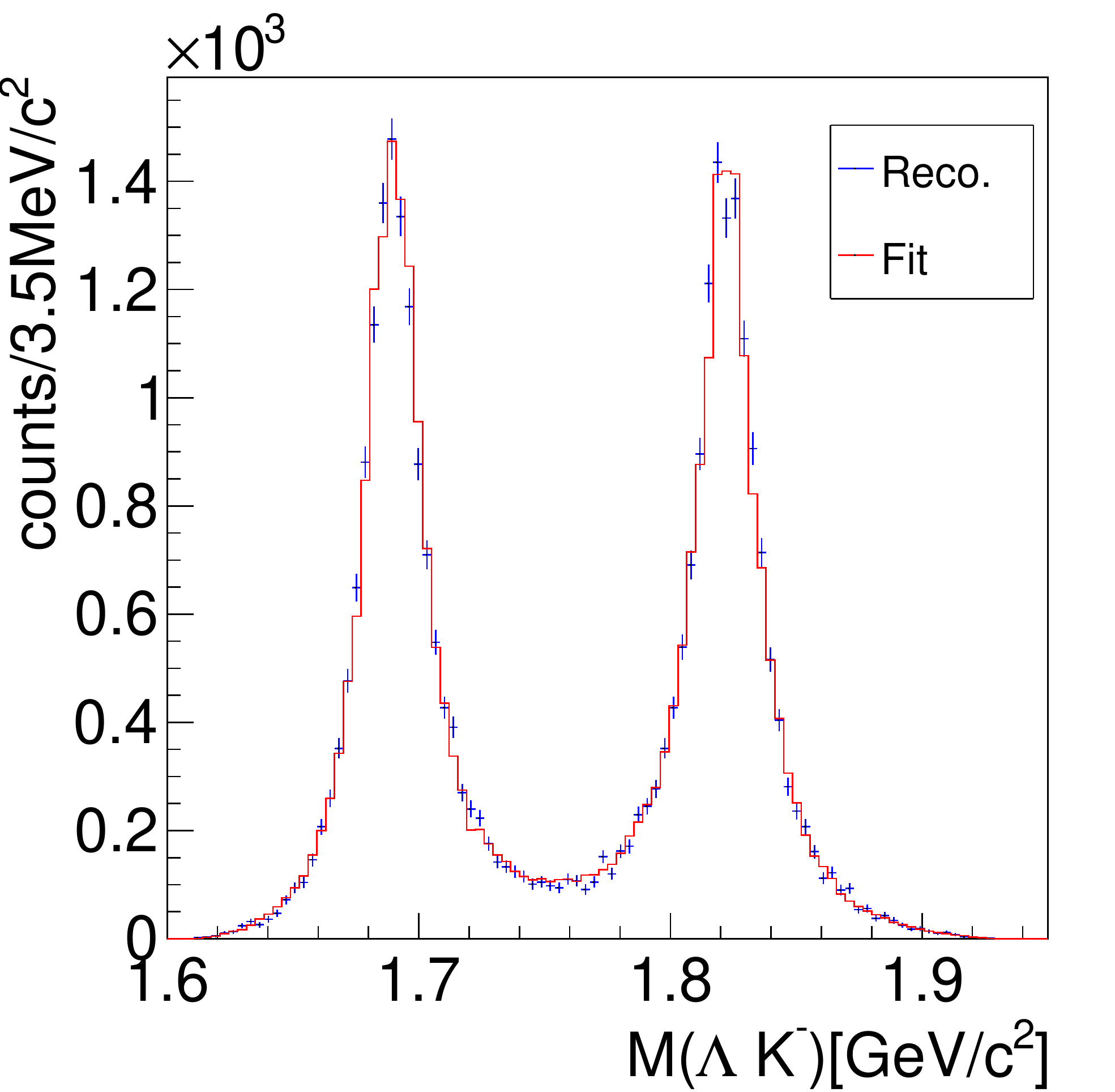}
	\caption{Comparison of the reconstructed (Reco.) and fitted $\Lambda$\kminus invariant mass distribution.}
	\label{fig:CombinedSampleInvariantMass}
\end{figure}
This can also be observed by comparing the invariant mass distributions for the generated sample and the best fit as illustrated in \cref{fig:CombinedSampleInvariantMass}.
The distribution for the fitted sample has been scaled by a factor $1/3$ to match the range of the generated distribution.\\
As described in \cref{sec:evtGen}, the dynamics of the process are modeled using a Breit-Wigner function. 
We are aware that in reality such a model might be too naive and that one would expect that a more complex description will be required including effects coming from coupled-channel phenomena. This is in particularly important when one aims to extract basic resonance parameters such as the mass and width of hyperon states.
We therefore concentrate primarily on the determination of the spin and parity of the specific resonances from PANDA data for which a modeling of the dynamics using a Breit-Wigner description is sufficient as a proof-of-principle for this single channel partial wave analysis.
\\
The event generator provided in PAWIAN generates the angular distribution of the \anticascade\excitedcascade system as well as of the \lam\kminus system according to the chosen spin and parity \gls{QN}.
The reconstructed as well as the fitted angular distribution for \lam in the helicity frame of the \excitedcascade is shown in \cref{fig:AngularDistrGenRecoLamHeli}.
The fitted distribution is in good agreement with the reconstructed one used as input for the fit.


\section{Conclusion $\mathbf{\&}$ Outlook}
	\label{sec:summary}
	First results for a feasibility study to determine the spin and parity \gls{QN} for specific $\Xi$ resonances states, \excitedcascadesixteen and \excitedcascadetwenty, have been presented.
The \gls{QN} are determined by performing a \gls{pwa} with the PAWIAN software.
The outcome of this study is promising and indicate, that the determination of the spin and parity quantum numbers of the $\Xi$ resonances from \panda data will be possible.
According to the feasibility study of the \fs final state \cite{Puetz2021}, it is expected that \panda is able to reconstruct between \SI{3}{\percent} and \SI{5}{\percent} of the generated events depending on the track reconstruction efficiency.
Assuming a luminosity $\mathcal{L} = 10^{31}\si{\per\centi\meter\squared\per\second}$ for the initial measurements at \panda, a signal reconstruction efficiency of \SI{3}{\percent} corresponds to about 10,500 reconstructed events per day.
This will give the possibility to collect the needed statistics for this \gls{pwa} within a few days of data taking at \panda \cite{Puetz2021} during the first phase of the experiment.\\
For this study, different scenarios were investigated, starting with the simplest case where the \pbarp reaction produces a single resonance and a \anticascade.
This simplified scenario was used to study the influence of the chosen \gls{QN} hypothesis on the reconstruction efficiency, the detector acceptance and detector resolution.
The chosen spin and parity values for the different data samples were shown to have no significant impact on the reconstruction efficiency of the data samples.
The mass resolution shows only slight deviations for the different hypotheses and in all cases the full angular range can be reconstructed.
Furthermore, the input \gls{QN} has been successfully reproduced with the multi-dimensional fit.\\
The study was then extended to investigate a data sample including both $\Xi$ resonances. 
One generated hypothesis has been tested so far: $J^P\left(\Xi\left(1690\right)\right),J^P\left(\Xi\left(1820\right)\right) = 1/2^+, 3/2^-$.
16 spin and parity combinations have been used for the multi-dimensional fit. 
The best fit result is achieved by using the true hypothesis and all wrong hypotheses can be excluded with high significance.
From \cite{Puetz2021} a good suppression of hadronic background is expected due to the kinematics of the decay chain.
Only a small fraction of wrongly combined events remains in the data sample after the analysis.
Therefore, possible background contributions are expected to have little influence on the fit results.\\
%
The models used in this study are likely a limited representation of reality.
In a next step, it would be appropriate to study a wider spectrum of models that include different models to describe the reaction dynamics, final state interactions and resonances in the \anticascade\kminus system.
Furthermore, a blind analysis of future \panda data should be performed to verify the results obtained in this study.

\section*{Acknowledgements}

We acknowledge financial support from
  the Bhabha Atomic Research Centre (BARC) and the Indian Institute of Technology Bombay, India;
  the Bundesministerium f\"ur Bildung und Forschung (BMBF), Germany;
  the Carl-Zeiss-Stiftung 21-0563-2.8/122/1 and 21-0563-2.8/131/1, Mainz, Germany;
  the Center for Advanced Radiation Technology (KVI-CART), Groningen, Netherlands;
  the CNRS/IN2P3 and the Universit\'{e} Paris-Sud, France;
  the CU (Czech Republic): MSMT LM2018112, OP VVV CZ.02.1.01/0.0/0.0/18\_046/0016066;
  the Deutsche Forschungsgemeinschaft (DFG), Germany;
  the Deutscher Akademischer Austauschdienst (DAAD), Germany;
  the European Union's Horizon 2020 research and innovation programme under grant agreement No 824093;
  the Forschungszentrum J\"ulich, Germany;
  the Gesellschaft f\"ur Schwerionenforschung GmbH (GSI), Darmstadt, Germany;
  the Helmholtz-Gemeinschaft Deutscher Forschungszentren (HGF), Germany;
  the INTAS, European Commission funding;
  the Institute of High Energy Physics (IHEP) and the Chinese Academy of Sciences, Beijing, China;
  the Istituto Nazionale di Fisica Nucleare (INFN), Italy;
  the Ministerio de EducaciÃ³n y Ciencia (MEC) under grant FPA2006-12120-C03-02, Spain;
  the Polish Ministry of Science and Higher Education (MNiSW) grant No. 2593/7, PR UE/2012/2, and the National Science Centre (NCN) DEC-2013/09/N/ST2/02180, Poland;
  the State Atomic Energy Corporation Rosatom, National Research Center Kurchatov Institute, Russia;
  the Schweizerischer Nationalfonds zur F\"orderung der Wissenschaftlichen Forschung (SNF), Switzerland;
  the Science and Technology Facilities Council (STFC), British funding agency, Great Britain;
  the Scientific and Technological Research Council of Turkey (TUBITAK) under the Grant No. 119F094, Turkey;
  the Stefan Meyer Institut f\"ur Subatomare Physik and the \"Osterreichische Akademie der Wissenschaften, Wien, Austria;
  the Swedish Research Council and the Knut and Alice Wallenberg Foundation, Sweden;
  the U.S. Department of Energy, Office of Science, Office of Nuclear Physics.

\newpage
\bibliographystyle{IEEEtran}
\bibliography{literature}




\end{document}